\documentstyle[aps,multicol,psfig]{revtex}
\begin{document}
\input {psfig}
\draft
\title{Ion-Induced Surface Diffusion in Ion Sputtering}

\author{Maxim  A. Makeev  and Albert-L\'aszl\'o Barab\'asi}

\address{Department of Physics, University of Notre Dame, Notre
Dame,
IN 46556}

\date{\today}

\maketitle

\begin{abstract}
Ion bombardment is known to enhance surface diffusion and affect the
surface morphology.  To quantify this phenomenon we calculate the
ion-induced diffusion constant and its dependence on the ion energy,
flux and angle of incidence. We find that ion bombardment can both
enhance and suppress diffusion and that the sign of the diffusion
constant depends on the experimental parameters.  The effect of
ion-induced diffusion on ripple formation and roughening of
ion-sputtered surfaces is discussed and summarized in a morphological
phase diagram.
\end{abstract}




\begin{multicols}{2}

\narrowtext

Sputtering, the removal of atoms from the surface of solids through
the impact of energetic particles (ions), is an important thin film
processing technique \cite{revsput}.  Consequently, much attention has
been focused on the measurement and calculation of the sputtering
yield and of  the  velocity and angular
distribution of the sputtered particles. However, for many
applications an equally important phenomenon, ion-induced surface
diffusion, has eluded sufficient understanding so far.
In the absence of ion bombardment surface diffusion is thermally
activated and characterized by the diffusion constant $D_T = D_0
\exp\left[- {E_d / k_B T}\right]$ such that the evolution of the
surface height $h(x,y,t)$ is described by the continuum equation
${\partial h / \partial t} = - D_T \nabla^4 h$ \cite{herr}. Here $E_d$
is the activation energy for surface diffusion of the adatoms and $T$
is the substrate temperature.  However, numerous experiments
investigated the effect of ion bombardment on island formation,
surface migration, surface smoothing and ripple formation have
provided evidence that ion bombardment is accompanied by an increase
in surface diffusion \cite{r2,r1,maclaren,r3,r4,rossn}.  In
particular, it has been demonstrated that ion-induced surface
diffusion can decrease the epitaxial temperature \cite{r2}, enhance
nucleation during growth \cite{r1}, or modify the surface morphology.
For example, MacLaren {\it et al.} \cite{maclaren} bombarded GaAs with
17 KeV Cs$^+$ in the temperature range from $-$50 to 200 $^\circ$C,
observing the development of a ripple structure on the surface with a
wavelength proportional to the square root of the diffusion constant.
When decreasing the temperature, the ripple spacing (wavelength) did
not decrease exponentially to zero with the inverse temperature, but
at approximately 60 $^\circ$C it stabilized at a constant value,
providing direct evidence for a temperature independent ion-induced
surface diffusion constant.

  Although the effect of the ions on surface diffusion is well
documented experimentally, there is no theory that would quantify it.
In this paper we calculate analytically the ion-induced diffusion
constant, $D^I$, and its dependence on the ion energy, flux, angle of
incidence, and penetration depth.  We find that there exists a
parameter range when ion bombardment generates a {\it negative}
surface diffusion constant, leading to morphological instabilities
along the surface, affecting the surface roughness and the ripple
structure.  The effect of ion-induced diffusion on the morphology of
ion-sputtered surfaces is summarized in a morphological phase diagram,
allowing for direct experimental verification of our predictions.

\begin{figure}[htb]
\begin{center}
\mbox{\psfig{figure=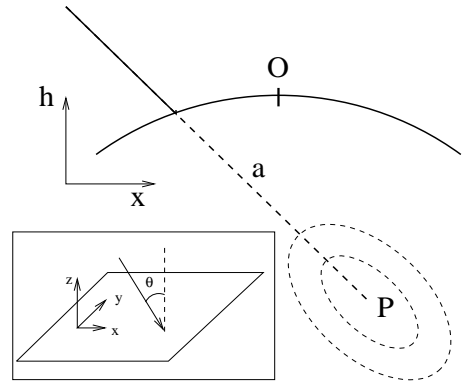,width=6cm,height=5 cm,angle=-90}}
\end{center}
\caption{Following a straight trajectory (solid line) the ion
penetrates an average distance $a$ inside the solid (dotted line)
after which it completely spreads out its kinetic energy. The energy
decreases with the distance from $P$, the dotted curves indicating
schematically the equal energy contours. The energy released at point
$P$ contributes to erosion at $O$. The inset shows the laboratory
coordinate frame: the ion beam forms an angle $\theta$ with the normal
to the {\it average} surface orientation, $z$, and the in-plane
direction $x$ is chosen along the projection of the ion beam. }
\label{fig1}
\end{figure}

  Ion-sputtering is determined by atomic processes taking place within a
finite penetration depth inside the bombarded material. The ions
penetrate the surface and transfer their kinetic energy to the atoms
of the substrate by inducing cascades of collisions among the
substrate atoms, or through other processes, such as electronic
excitations. A convenient picture of the ion bombardment process is
shown in Fig. \ref{fig1}. The ions penetrate a distance $a$ inside
the solid before they completely spread out their kinetic energy with
some assumed spatial distribution.  An ion releasing its energy at
point $P$ in the solid contributes energy to the surface point $O$,
that may induce the atoms in $O$ to break their bonds and either leave
the surface or diffuse along it.

  Following \cite{sig,bh}, we consider
that the average energy deposited at point $O$ due to the ion arriving
at $P$ follows the Gaussian distribution
\begin{equation}
E(\mbox{\boldmath $r'$})  =  \frac{\epsilon}{(2\pi)^{3/2} \sigma
\mu^2} \exp \left\{ - \frac{z'^2}{2 \sigma^2} -
\frac{x'^2+y'^2}{2 \mu^2} \right\}.
\label{gaussian}
\end{equation}

In (\ref{gaussian}) $z'$ is the distance measured along the ion
trajectory, $x'$, $y'$ are measured in the plane perpendicular to it;
$\epsilon$ denotes the kinetic energy of the ion and $\sigma$ and
$\mu$ are the widths of the distribution in directions parallel and
perpendicular to the incoming beam, respectively.  However, the sample
is subject to an uniform flux $J$ of bombarding ions. A large number
of ions penetrate the solid at different points simultaneously and the
velocity of erosion at $O$ depends on the total power ${\cal E}_O$
contributed by all the ions deposited within the range ${\cal R}$ of
the distribution (\ref{gaussian}), such that

\begin{equation}
v = p \; \int_{{\cal R}} d\mbox{\boldmath $r$} \;
\Phi(\mbox{\boldmath $r$})
\; E(\mbox{\boldmath $r$}),
\label{vel}
\end{equation}
where $\Phi(\mbox{\boldmath $r$})$ corrects for the local slope
dependence of the uniform flux $J$ and $p$ is a proportionality
constant between power deposition and the erosion rate \cite{sig}.
The calculation of $v$ involves the following assumptions (for more
details see Refs. \cite{bh,us}): (a) In the {\em laboratory}
coordinate frame $(x,y,z)$ the surface can be described by a single
valued height function $h(x,y,t)$, measured from an initial flat
configuration which lies in the ($x$,$y$) plane (see Fig. \ref{fig1});
(b) The angle between the ion beam direction and the local normal to
the surface is a function of the angle of incidence $\theta$ and the
values of the local slopes $\partial_x h$ and $\partial_y h$, and can
be expanded in powers of the latter.  Under these conditions, we can
expand (\ref{vel}), obtaining the equation of motion \cite{Comment}

\begin{eqnarray}
\frac{\partial h}{\partial t} & = & -v_0 + \gamma
\frac{\partial h}{\partial x} + \nu_x \frac{\partial^2
h}{\partial
x^2} +
\nu_y \frac{\partial^2 h}{\partial y^2} + \nonumber \\
& & + \frac{\lambda_x}{2} \left(
\frac{\partial h}{\partial x} \right)^2 + \frac{\lambda_y}{2}
\left(
\frac{\partial h}{\partial y} \right)^2 - D^I_x \frac{\partial^4
h}{\partial x^4}
- D^I_y \frac{\partial^4 h}{\partial y^4}.
\label{eqn}
\end{eqnarray}

\begin{figure}[htb]
\begin{center}
\hskip 3.0 cm 
\mbox{\psfig{figure=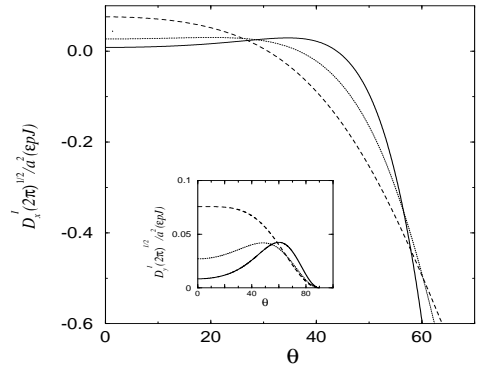,width=6.5 cm,height=5 cm,angle=-90}}
\end{center}
\caption{Ion-induced diffusion constant, $D_x^I$ and $D_y^I$ (inset)
as a function of the angle of incidence $\theta$. In both figures the
curves correspond to $a_\sigma=1.0$ (dashed line), $a_\sigma=1.5$
(dotted line) and $a_\sigma=2.0$ (continuous line).}
\label{fig2}
\end{figure}

 From (\ref{vel}) we can calculate the expressions for the
coefficients appearing in (\ref{eqn}) in terms of the physical
parameters which characterize the sputtering process. The coefficients
$\nu_x$ and $\nu_y$ were first calculated by Bradley and Harper
\cite{bh}, while the nonlinear expansion was performed in \cite{us},
providing $\lambda_x$ and $\lambda_y$.  A fourth order expansion is
used to obtain the ion-induced diffusion constants $D^I_x$ and
$D^I_y$.  To simplify the discussion we restrict ourselves to the
symmetric case $\sigma=\mu$.  Using $F \equiv (\epsilon J
p/\sqrt{2\pi}) \exp(-a_{\sigma}^2/2+a_{\sigma}^2 s^2) $, $s \equiv
\sin \theta$, $c \equiv \cos \theta$ and $a_{\sigma} \equiv a/\sigma$,
we find for the surface diffusion constants
\begin{equation}
D^I_x = \frac{F a^{2}}{24 a_{\sigma} } \left\{a_{\sigma}^4 s^4 c^2 +
 a_{\sigma}^2 ( 6 c^2 s^2 -4 s^4 ) + 3 c^2 -12 s^2 \right\},
\label{Dx}
\end{equation}
\begin{equation}
D^I_y = \frac{F a^{2}}{24 a_{\sigma} } 3 c^2.
\label{Dy}
\end{equation}
Consistent with symmetry considerations for $\theta=0$ we obtain
$D^I_x=D^I_y$. However, for $\theta \neq 0$ we find that $D^I_x \neq
D^I_y$, i.e. the ion-induced surface diffusion is {\it anisotropic}.
Moreover, its sign also depends on the experimental parameters.  The
consequences of (\ref{Dx}) and (\ref{Dy}) can be summarized as
follows: (a) Independent of the angle of incidence $D^I_y$ is positive
while the sign of the $D^I_x$ depends on both $\theta$ and $a_\sigma$
as shown in Fig. \ref{fig2}.

\vskip 0.2 cm 

  Thus, while for $\theta=0$ the ion
bombardment enhances the surface diffusion ($D^I_{x} > 0$), for large
$\theta$ it can suppress diffusion; (b) The fact that $D_x^I$ can be
negative indicates that any simple theory connecting the magnitude of
the ion-induced diffusion to the energy transferred by the ions to the
surface is incomplete, since it can predict only a positive $D^I$. In
fact, $D^I$ is the result of a complex interplay between the local
{\it surface topology} and the energy transferred to the surface; (c)
The diffusion constants are proportional to the flux $J$, in agreement
with the detailed experimental study of Cavaille and Dreschner
\cite{r4}; (d) It is a standard experimental practice to report the
magnitude of the ion-enhanced diffusion using an effective temperature
$T^{eff}$ at which the substrate needs to be heated to obtain the same
mobility as with ion bombardment \cite{r4,rossn}.  We can calculate
$T^{eff}$ using the relation $D^I+ D_0 \exp(-E_a/k_BT) = D_0 \exp(-E_a
/k_B T^{eff})$, that has two important consequences.  First, the
anisotropic diffusion constant translates into an anisotropic
$T^{eff}$, i.e. we have $T^{eff}_x \neq T^{eff}_y$. The experimental
methods used to estimate $T^{eff}$ could not distinguish $T^{eff}_x$
and $T^{eff}_y$ \cite{r4,rossn}. However, current observational
techniques should be able to detect the difference between the two
directions.  Second, while it is generally believed that ion
bombardment can only raise the effective temperature since it
transfers energy to the surface, the negative $D^I$ indicates that
along the $x$ direction one could have $T_{eff} < T$. (e) Finally, the
results (\ref{Dx})-(\ref{Dy}) are based on Sigmund's theory of
sputtering \cite{sig} that describes sputtering in the linear cascade
regime. The energy range when this approach is applicable lies between
0.5 KeV and 1 MeV, the precise lower and upper limits being material
dependent.  Thus, we do not expect (\ref{Dx})-(\ref{Dy}) to apply to
low energy (few eV) ion-enhanced epitaxy.

{\it Ripple formation---} Sputtering may lead to the development of a
 ripple morphology on the surface \cite{chason}.  The origin of the
 ripple formation is an ion-induced instability: valleys are eroded
 faster than crests, expressed by negative $\nu_x$ and $\nu_y$
 coefficients in (\ref{eqn}) \cite{bh}. At short wavelength this
 instability is balanced by surface diffusion.  A linear stability
 analysis predicts that the observable ripple wavelength is determined
 by the most unstable mode, and has the wavelength $\ell = 2 \pi
 \sqrt{D/|\nu|}$, where $\nu$ is the largest in absolute value of the
 negative surface tension coefficients. Accordingly, the wave vector
 of the ripples is parallel to the $x$ axis for small $\theta$ and
 perpendicular to it for large $\theta$ \cite{bh}. The large length
 scale behavior is described by the noisy anisotropic
 Kuramoto-Sivashinsky equation \cite{us,ks}, leading to roughening
 \cite{eklund} or the development of coarsening ripple domains 
 \cite{rost}.

{\it Quantitative comparison with experiments---} At finite
temperature the total diffusion constant given by $D= D^I +
D_T$. As T decreases there is a critical temperature, $T_c$, at which
$D^I = D_T$, so that for $T<T_c$ the diffusion is dominated by its
ion-induced component, which is independent of temperature, in
agreement with the experimental results of MacLaren {\it et al.}
\cite{maclaren}.  Unfortunately, for most materials the quantities
entering in $F$, $a_\sigma$ and $D_0$ are either unknown, or only
their order of magnitude can be estimated. However, using the
expression for $\ell$ we can express $T_c$ in terms of measurable
quantities free of these constants
\begin{equation}
T_c={T_0 \over 1-(2 k_B  T_0/E_a) \ln (\ell_{Ion}/\ell_{T_0})}
\label{tc}
\end{equation}
where $\ell_{T_0}$ is the experimentally measured ripple wavelength at
any temperature $T_0> T_c$; $\ell_{Ion}$ is the ripple wavelength in
the low temperature regime, $T < T_c$, where ion induced diffusion
dominates, and therefore $\ell_{Ion}$ is independent of $T$; $E_a$ is
provided by the slope of $\ln(\ell)$ versus $1/T$ in the high
temperature regime ($T >> T_c$). Consequently, {\it all} quantities in
(\ref{tc}) can be obtained from a plot of the ripple wavelength as a
function of temperature, so that (\ref{tc}) gives $T_c$ {\it without
any free parameters}. Such a plot is provided by MacLaren {\it et al.}
\cite{maclaren}, leading to $E_a = 0.51$eV, $\ell_{Ion}=0.8\mu$m.
Using $\ell_{T_0}=2\mu$m for $T=368$K \cite{maclaren}, we obtain
$T_c=57^o$C, which is in good agreement with the experiments, that
provide $T_c$ between 45 and 60$^o$C \cite{maclaren}.

\begin{figure}[htb]
\begin{center}
\hskip 2.0 cm
\psfig{figure=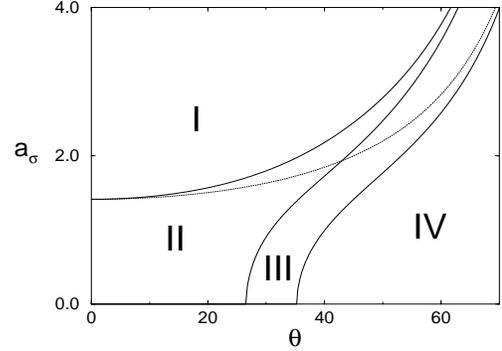,width=7 cm,height=5 cm,angle=-90}
\end{center}
\vskip 0.1 cm
\caption{Phase diagram for the isotropic case $\sigma=\mu=1$.  Region
I: $\nu_x<0$, $\nu_y < 0$, $D_x^I >0$, $D_y^I > 0$ and $\ell_x >
\ell_y$; Region II: $\nu_x<0$, $\nu_y < 0$, $D_x^I >0$, $D_y^I > 0$,
and $\ell_x < \ell_y$; Region III: $\nu_x<0$, $\nu_y < 0$, $D_x^I <0$
and $D_y^I > 0$; Region IV: $\nu_x>0$, $\nu_y < 0$, $D_x^I <0$ and
$D_y^I > 0$. Note that the phase diagram is independent of the precise
values of $J$,$p$, while the $\epsilon$ dependence is contained in
$a_\sigma$. }
\label{fig3}
\end{figure}

{\it Morphological phase diagram---} The detailed morphological phase
diagram is rather complex if the diffusion is not thermally activated
but ion-induced. At low temperatures, when $D_T$ is negligible
compared to $D^I$, the ripple wavelengths are $\ell_x^I = 2 \pi
\sqrt{D^I_x/|\nu_x|}$ and $\ell_y^I = 2 \pi \sqrt{D^I_y/|\nu_y|}$. In
the following we discuss the dependence of the surface morphologies on
the experimental parameters $\theta$ and $a_\sigma$, based on the
phase diagram shown in Fig.  \ref{fig3}.

  {\it Region I---} The surface tensions, $\nu_x$ and $\nu_y$, are
negative while $D_x$ and $D_y$ are positive, consequently we have a
superimposed ripple structure along the $x$ and the $y$ directions.
The experimentally observed ripple wavelength is the smallest of the
two, and since $\ell^I_x > \ell^I_y$, the ripple wave vector is
oriented along the $x$ direction.  The lower boundary of this region
separating it from Region II is given by the solution of the
$\ell^I_x=\ell^I_y$ equation.

  {\it Region II---} Here the ripple wave vector is oriented along the
$y$ direction, since $\ell^I_x < \ell^I_y$. This region is bounded
below by the $D^I_{x}=0$ line.  At large length scales in I and II one
expects kinetic roughening described by the KPZ equation
\cite{us,revrough,kpz}.

  {\it Region III ---} In this region $D^I_{x}$ is negative, while the
sign of all other coefficients are the same as in I and II.  Since
both surface tension and surface diffusion are destabilizing along
$x$, every mode is unstable and one expects that the KPZ nonlinearity
cannot turn on the KS stabilization \cite{us}, the system being
unstable at large length scales as well. This instability is expected
to lead to exponential roughening. The lower boundary of this region
is given by the $\nu_x=0$ line.

 {\it Region IV ---} Here we have $\nu_x > 0$, $\nu_y<0$, $D_{xx}^I <
0$ and $D_{yy}^I > 0$, i.e. one expects the surface to be periodically
modulated in the $y$ direction, leading to a ripple structure
oriented along the $x$ direction.  In the $x$ direction we have a
reversal of the instability: the short length scale instability
generated by the negative $D^I_x$ is stabilized by the positive
surface tension $\nu_x$. Thus, there is no ripple structure along the
$y$ direction. Regarding the large length scale behavior, along the
$x$ direction the surface diffusion term is irrelevant compared to the
surface tension, thus one expects KPZ scaling.  However, along the $y$
direction the KS mechanism is expected to act, renormalizing the
negative $\nu_y$ to positive values for length scales larger than
$\ell^I_y$, leading to a large wavelength KPZ behavior.

If thermal and ion-induced diffusion coexist, the ripple wavelengths
are given by $\ell_x=2 \pi [(D_T + D_{x}^I)/|\nu_x|]^{1/2}$ and
$\ell_y=2 \pi [(D_T + D_{y}^I)/|\nu_y|]^{1/2}$. The phase diagram for
intermediate temperatures can be calculated using the total $D$.  In
particular for high $T$, when $D_T >> D_{x}^I$ and $D_T >> D_{y}^I$,
the phase diagram converges to the one obtained in Ref. \cite{us}, the
ripple orientation being controlled by the $\nu_x=\nu_y$ line (dotted
in Fig. \ref{fig3}).  Thus,  with increasing $D_T$ the phase boundary
between the regions I and II converges to the $\nu_x=\nu_y$ line and
the $D_{x}=0$ boundary separating the regions II and III shifts
downwards, eventually disappearing. However, in the intermediate
regions new phases with coarsening ripple domains \cite{rost} appear
as the $D^I_{x}$ and the $\nu_x=0$ lines cross each other.

  While we limited our discussion to the effect of the ion-induced
diffusion on the surface morphology, the results (\ref{Dx})-(\ref{Dy})
can be used to investigate other phenomena as well, such as island
nucleation. The experimental verification of the above results would
constitute an important step to elucidate the mechanism responsible
for ion-induced diffusion, with potential applications to ion-enhanced
epitaxy.

  We would like to acknowledge discussions and comments by R. Cuerno. 
This research was partially supported by the University of Notre Dame
Faculty Research Program.

\end{multicols}


\begin{thebibliography}{99}
\bibitem{Comment} In the expansion of (\ref{gaussian}) additional
second, third and fourth order (linear and nonlinear) terms are
obtained on the rhs of Eqn. (\ref{eqn}) that are not shown here, since
they are irrelevant regarding the discussed phenomena (see also
Ref. \cite{us}).
\end{thebibliography}
\end{document}